\documentclass[12pt]{article}
\usepackage{amsfonts,amssymb,amsmath}
\usepackage[dvips]{epsfig}
\begin{document}

\def\ds{\displaystyle}
\def\beq{\begin{equation}}
\def\eeq{\end{equation}}
\def\bea{\begin{eqnarray}}
\def\eea{\end{eqnarray}}
\def\beeq{\begin{eqnarray}}
\def\eeeq{\end{eqnarray}}
\def\ve{\vert}
\def\vel{\left|}
\def\bpll{B\rar\pi \ell^+ \ell^-}
\def\ver{\right|}
\def\nnb{\nonumber}
\def\ga{\left(}
\def\dr{\right)}
\def\aga{\left\{}
\def\adr{\right\}}
\def\lla{\left<}
\def\rra{\right>}
\def\rar{\rightarrow}
\def\nnb{\nonumber}
\def\la{\langle}
\def\ra{\rangle}
\def\ba{\begin{array}}
\def\ea{\end{array}}
\def\tr{\mbox{Tr}}
\def\ssp{{\Sigma^{*+}}}
\def\sso{{\Sigma^{*0}}}
\def\ssm{{\Sigma^{*-}}}
\def\xis0{{\Xi^{*0}}}
\def\xism{{\Xi^{*-}}}
\def\qs{\la \bar s s \ra}
\def\qu{\la \bar u u \ra}
\def\qd{\la \bar d d \ra}
\def\qq{\la \bar q q \ra}
\def\gGgG{\la g^2 G^2 \ra}
\def\q{\gamma_5 \not\!q}
\def\x{\gamma_5 \not\!x}
\def\g5{\gamma_5}
\def\sb{S_Q^{cf}}
\def\sd{S_d^{be}}
\def\su{S_u^{ad}}
\def\ss{S_s^{??}}
\def\ll{\Lambda}
\def\lb{\Lambda_b}
\def\sbp{{S}_Q^{'cf}}
\def\sdp{{S}_d^{'be}}
\def\sup{{S}_u^{'ad}}
\def\ssp{{S}_s^{'??}}
\def\sig{\sigma_{\mu \nu} \gamma_5 p^\mu q^\nu}
\def\fo{f_0(\frac{s_0}{M^2})}
\def\ffi{f_1(\frac{s_0}{M^2})}
\def\fii{f_2(\frac{s_0}{M^2})}
\def\O{{\cal O}}
\def\sl{{\Sigma^0 \Lambda}}
\def\es{\!\!\! &=& \!\!\!}
\def\ar{&+& \!\!\!}
\def\ek{&-& \!\!\!}
\def\cp{&\times& \!\!\!}
\def\se{\!\!\! &\simeq& \!\!\!}
\def\hml{\hat{m}_{\ell}}
\def\rr{\hat{r}_{\Lambda}}
\def\ss{\hat{s}}

\renewcommand{\textfraction}{0.2}    
\renewcommand{\topfraction}{0.8}
\renewcommand{\bottomfraction}{0.4}
\renewcommand{\floatpagefraction}{0.8}
\newcommand\mysection{\setcounter{equation}{0}\section}

\def\baeq{\begin{appeq}}     \def\eaeq{\end{appeq}}
\def\baeeq{\begin{appeeq}}   \def\eaeeq{\end{appeeq}}
\newenvironment{appeq}{\beq}{\eeq}
\newenvironment{appeeq}{\beeq}{\eeeq}
\def\bAPP#1#2{
 \markright{APPENDIX #1}
 \addcontentsline{toc}{section}{Appendix #1: #2}
 \medskip
 \medskip
 \begin{center}      {\bf\LARGE Appendix #1 :}{\quad\Large\bf #2}
\end{center}
 \renewcommand{\thesection}{#1.\arabic{section}}
\setcounter{equation}{0}
        \renewcommand{\thehran}{#1.\arabic{hran}}
\renewenvironment{appeq}
  {  \renewcommand{\theequation}{#1.\arabic{equation}}
     \beq
  }{\eeq}
\renewenvironment{appeeq}
  {  \renewcommand{\theequation}{#1.\arabic{equation}}
     \beeq
  }{\eeeq}
\nopagebreak \noindent}

\def\eAPP{\renewcommand{\thehran}{\thesection.\arabic{hran}}}
\renewcommand{\theequation}{\arabic{equation}}
\newcounter{hran}
\renewcommand{\thehran}{\thesection.\arabic{hran}}
\def\bmini{\setcounter{hran}{\value{equation}}
\refstepcounter{hran}\setcounter{equation}{0}
\renewcommand{\theequation}{\thehran\alph{equation}}\begin{eqnarray}}
\def\bminiG#1{\setcounter{hran}{\value{equation}}
\refstepcounter{hran}\setcounter{equation}{-1}
\renewcommand{\theequation}{\thehran\alph{equation}}
\refstepcounter{equation}\label{#1}\begin{eqnarray}}


\newskip\humongous \humongous=0pt plus 1000pt minus 1000pt
\def\caja{\mathsurround=0pt}

\title{\bf Multi-Dimensional Cosmology and DSR-GUP}
\author{K. Zeynali$^{1,2}$\thanks{Email: k.zeinali@arums.ac.ir}
\hspace{2mm},  \hspace{2mm}
 F. Darabi$^3$\thanks{Email: f.darabi@azaruniv.edu (Corresponding author)   }
 \hspace{2mm}, and \hspace{2mm}
 H. Motavalli$^2$\thanks{Email: motavalli@tabrizu.ac.ir } \\
\centerline{$^1$\small {\em Faculty of Medicine, Ardabil University of Medical Sciences (ArUMS), Daneshgah St., Ardabil, Iran.}}\\
\centerline{$^2$\small {\em Department of Theoretical Physics and
Astrophysics, University of Tabriz, 51666-16471, Tabriz,
Iran.}}\\
{$^3$\small {\em Department of Physics, Azarbaijan Shahid Madani University , 53714-161, Tabriz, Iran. }}}

\maketitle
\begin{abstract}
A multidimensional cosmology with FRW type metric having 4-dimensional space-time and $d$-dimensional Ricci-flat internal space is considered with a higher dimensional cosmological constant. The classical cosmology in commutative and DSR-GUP contexts is studied and the corresponding exact solutions for negative and positive cosmological constants are obtained. In the positive
cosmological constant case, it is shown that unlike the commutative as well
as GUP cases, in DSR-GUP case both scale factors of internal and external
spaces after accelerating phase will inevitably experience decelerating
phase leading simultaneously to a big crunch. This demarcation from GUP originates from the difference between the GUP and DSR-GUP algebras. The important result is that unlike GUP which results in eternal acceleration, DSR-GUP at first
generates acceleration but prevents the eternal acceleration at late times and turns it into deceleration.

\end{abstract}

~~~PACS numbers: 98.80.Hw; 04.50.+h

\section{Introduction}

The Generalized Uncertainty Principal (GUP) is a generalization of
Heisenberg Uncertainty Principal in the Planck scale where the
gravitational effects on quantum gravity may be considerable. This
idea, was firstly considered by Mead \cite{Mead} and then implemented
in the context of string theory as a candidate of quantum gravity
as well as black hole physics with the prediction of a minimum
measurable length \cite{Gross,Kato,general
GUP3,string1,string3,string6,Kemp}. Doubly Special Relativity
(DSR) theory \cite{DSR} as a possible ingredient of the flat
space-time limit of the quantum theory of gravity proposed another
modification on Heisenberg Uncertainty Principal \cite{DSR4}.
Recently the authors in \cite{Ali} considered these two
modification as a limit of a single algebra (DSR-GUP).

Nowadays, a large amount of interest has been focused on the effects
of these modification on system in high energy physics
\cite{sepangi}. In a recent paper \cite{zeynali}, we have studied a multi-dimensional Cosmology with GUP and obtained the corresponding exact
solutions for negative and positive cosmological constants. Especially, for
positive cosmological constant, the solutions revealed late time accelerating
behavior and internal space stabilized to the sub-Planck size, in good agreement with current observations. Motivated by the interest in DSR-GUP, in the present paper we are interested in studying the effects of DSR-GUP modifications on our multi dimensional cosmology and comparing its results with the GUP case\footnote{Throughout the paper we will use the units $\hbar = G = c = 1$, where $G$ is the gravitational constant and $c$ is the velocity of light.}. 

In section 2, we introduce the notions of GUP and DSR-GUP as well as their corresponding algebras. In section 3, we briefly introduce our cosmological model.  In section 4, we first obtain the commutative solutions and then find the DSR-GUP solutions. Finally, in conclusion, we compare the commutative,
GUP and DSR-GUP solutions.

\section{Generalized uncertainty principal}

The simplest form of the GUP in a one dimensional system can be
written as \cite{Kemp}

\bea\label{e1}
 \delta x \delta p\geq\frac{\hbar}{2}\Bigg(1+\beta L_{Pl}^2(\delta p)^2 \Bigg),
 \eea
where $L_{Pl}\sim10^{-35}m$ is the Planck length and $\beta$ is
a constant of order unity. The algebra corresponding to (\ref{e1}) can be
written as \cite{Kemp}

\bea\label{e2}
  [x_i,p_j]=i \{\delta_{ij}+\beta{L^2_{Pl}}(p^2 \delta_{ij} +2p_i
  p_j)\},
 \eea
which reduces to the ordinary one for $\beta\rightarrow 0$. Doubly Special Relativity theories, on the other hand, suggest that the planck scales
similar to the light speed are observer independent scales. This is because different observers should not observe quantum gravity effects at different scales \cite{DSR}. The algebra corresponding to DSR-GUP
can be written as \cite{DSR4}
\bea\label{e3}
  [x_i,p_j]=i \{\delta_{ij}-L_{Pl}|\overrightarrow{p}| \delta_{ij}+L^2_{Pl}p_ip_j\},
 \eea
which reduces to the ordinary one for $L_{Pl}\rightarrow 0$. 
The authors in \cite{Ali} showed that by assumption
$[x_i,x_j]=0=[p_i,p_j]$, the two above algebra (\ref{e2}),
(\ref{e3}) can be considered as a single algebra in phase space

 \bea\label{e4}
  [x_i,p_j]=i \Bigg\{\delta_{ij}-a L_{Pl}\Bigg(p \delta_{ij}+
  \frac{p_i p_j}{p}\Bigg)+a^2 L^2_{Pl}(p^2 \delta_{ij} +3p_i
  p_j)\Bigg\},
 \eea
where $a$ is assumed to be of order unity and $p^2=\sum p_i p_i$.
By definition \cite{Ali}

\bea\label{e5}
 x_{i}=x_{i0}, &  & p_i=p_{i0}(1-a L_{Pl} p_0+2a^2 L^2_{Pl}p^2_0),
 \eea
the equation (\ref{e4}) can be satisfied, where $x_{i0}$ and $p_{i0}$
are the ordinary position and momenta with $[x_{i0},p_{j0}]=i
\delta_{ij}$ and $p_{j0}=-i\frac{\partial}{\partial x_{i0}}$. To
distinguish between the linear and second order terms in Planck
length, we rewrite equation (\ref{e5}) in a more general form

\bea\label{e6}
 x_{i}=x_{i0}, &  & p_i=p_{i0}(1-\alpha L_{Pl} p_0+\beta L^2_{Pl}p^2_0).
 \eea
Here, the coefficient $\alpha$ indicates the effect of linear term
in Planck length and $\beta$ the effect of second order term in
Planck length. So setting $\alpha=0$ and $\alpha=a,\beta=2a^2$
gives back the ordinary GUP algebra (\ref{e2}) and the DSR-GUP
algebra (\ref{e4}), respectively. Using (\ref{e6}), we can show
that the $p^2$ term in the any Hamiltonian can be be derived as

\bea\label{e7}
 p^2=p_0^2-2\alpha L_{Pl} p_0^3+(\alpha^2+2\beta) L^2_{Pl}p^4_0.
 \eea

\section{The Cosmological Model}

We consider a multi-dimensional cosmology in which the space-time is established by a FRW type metric with 4-dimensional space-time and a d-dimensional Ricci-flat internal space \cite{khosravi1}

\bea\label{e8}
ds^2=-dt^2+\frac{R^2(t)}{(1+\frac{k}{4}r^2)}(dr^2+r^2d\Omega^2)+a^2(t)g_{ij}^{(d)}dx^idx^j,
  \eea
where $R(t)$ and $a(t)$ are the scale factors of the external and
internal spaces respectively, and $g_{ij}^{(d)}$ is the Ricci-flat metric of the internal space. The Ricci scalar is derived from the metric (\ref{e8}) \cite{khosravi1}
  \bea\label{e9}
  {\cal R}=6\Bigg(\frac{\ddot{ R}}{ R}+\frac{k+\ddot{ R}^2}{R^2}\Bigg)
  +2d\frac{\ddot{a}}{a}+d(d-1)\Bigg(\frac{\dot{a}}{a}\Bigg)^2+6d\frac{\dot{a}\dot{R}}{a R},
  \eea
where a dot represents differentiation with respect to time $t$. 
The Einstein-Hilbert action with a $(3+d)$-dimensional cosmological constant $\Lambda$ is written as

  \bea\label{e10}
  {\cal S}=\frac{1}{2k^2_{3+d}}\int_M
  d^{4+d}x\sqrt{-g}({\cal R}-2\Lambda)+{\cal S}_{YGH},
  \eea
where $k_{3+d}$ is the $(3+d)$-dimensional gravitational constant
and ${\cal S}_{YGH}$ is the York-Gibbons-Hawking boundary term.
By substituting (\ref{e9}) in (\ref{e10}) and dimensional reduction we have

  \bea\label{e11}
  {\cal S}=-v_{3+d}\int dt\Bigg \{6\dot{ R}^2\Phi R+6\dot{ R}\dot{\Phi}
  R^2+\frac{d-1}{d}\frac{\dot{\Phi}^2}{\Phi} R^3-6k\Phi R+2\Phi R^3\Lambda \Bigg\},
  \eea
where
  \bea\label{e12}   \Phi=\Bigg(\frac{a}{a_0}\Bigg)^d,
  \eea
and $a_0$ is the present time compactification scale of the internal space. We introduce the following change of variables provided $v_{3+d}=1$ \cite{zeynali}
  \bea\label{e13}
   \Phi  R^3=\Upsilon^2(x_1^2-x_2^2),
  \eea

  \bea\label{e14}
   \Phi^{\rho_+}  R^{\sigma_-}&=&\Upsilon(x_1+x_2),\nnb \\
   \Phi^{\rho_-}  R^{\sigma_+}&=&\Upsilon(x_1-x_2).
  \eea
with
   \bea\label{e15}
   \rho_\pm
   &=&\frac{1}{2}\pm\frac{1}{2}\sqrt{\frac{3}{d(d+2)}},
\nnb \\
   \sigma_\pm
   &=&\frac{3}{2}\pm\frac{1}{2}\sqrt{\frac{3d}{d+2}},\nnb \\
   \Upsilon&=&\frac{1}{2}\sqrt{\frac{d+3}{d+2}},
  \eea
where $R=R(x_1, x_2)$ and $\Phi=\Phi(x_1, x_2)$ are functions of
new variables $x_1, x_2$. The above transformations with $k=0$ result in the Lagrangian and Hamiltonian as follows
  \bea\label{e16}
   {\cal L}=
   (\dot{x_1}^2-\dot{x_2}^2)+\frac{\Lambda}{2}\Bigg(\frac{d+3}{d+2}\Bigg)\Big(x_1^2-x_2^2\Big),
  \eea
  \bea\label{e17}
   {\cal H}=\left(\frac{p_1^2}{4}+\omega^2x_1^2\right)-\left(\frac{p_2^2}{4}+\omega^2
   x_2^2\right),
  \eea
where
  \bea\label{e18}
   \omega^2=-\frac{1}{2}\Bigg(\frac{d+3}{d+2}\Bigg)\Lambda.
  \eea

\section{Solutions}

\subsection{Commutative case}

  The dynamical variables defined in (\ref{e14}) and their conjugate momenta
 satisfy the following Poisson bracket algebra \cite{khosravi1, Matej Pavsic}
 \bea\label{e19}
   \{x_\mu, p_\nu \}_P=\eta_{\mu\nu},
  \eea
where $\eta_{\mu\nu}$ is the two dimensional Minkowski metric. The equations of motion are obtained 
 \bea\label{e21}
\ddot{x_\mu}+\omega^2 x_\mu=0.
 \eea
For a negative cosmological constant $\omega^2$ is positive and Eq.(\ref{e21}) describes the equations of motion for two ordinary uncoupled harmonic oscillators with solutions 

\bea\label{e22} x_\mu(t)=A_\mu e^{i\omega t}+B_\mu e^{-i\omega t},
 \eea
where $A_\mu$ and $B_\mu$ are constants of integration satisfying
$A_\mu B^\mu=0$ due to the Hamiltonian constraint $({\cal H}=0)$.
Using (\ref{e12}) and (\ref{e14}), the solutions for scale factors take
the following forms
 \bea\label{e24}
R(t)&=&k_2[\sin(\omega t+\phi_1)]^{\frac{-\rho_-}{\rho_+ \sigma_+
-\rho_- \sigma_-}}[\sin(\omega t+\phi_2)]^{\frac{\rho_+}{\rho_+
\sigma_+ -\rho_- \sigma_-}},\\
a(t)&=&k_1[\sin(\omega t+\phi_1)]^{\frac{\sigma_+}{d(\rho_+
\sigma_+ -\rho_- \sigma_-)}}[\sin(\omega
t+\phi_2)]^{\frac{-\sigma_-}{d(\rho_+ \sigma_+ -\rho_-
\sigma_-)}},
 \nnb 
 \eea
where $k_1$ and $k_2$ are arbitrary constants and $\phi_1$ and
$\phi_2$ are arbitrary phases. Imposing the Hamiltonian
constraint leads to the following relation 
 \bea\label{e25}
\frac{4(d+2)}{d+3}k_1^d k_2^3\cos(\phi_1-\phi_2)=0,
 \eea
where because of $k_1$, $k_2\neq0$, it results in
$\phi_1-\phi_2=\frac{\pi}{2}$. In what follows, we will investigate the
behavior of a Universe with one internal dimension $(D=3+1)$. By setting $\phi_1=\frac{\pi}{2}$ and $\phi_2=0$, we obtain
 \bea\label{e26}
R(t)&=&k_2\sqrt{\sin(\omega t)},\nnb \\
a(t)&=&k_1\frac{\cos(\omega t)}{\sqrt{\sin(\omega t)}}.
 \eea
The Hubble and deceleration parameters for both $R(t)$ and $a(t)$ are calculated
as
\bea\label{e28}
H_R(t)&=&\frac{\dot{R}(t)}{R(t)}=\frac{\omega}{2}\cot(\omega t),
\nnb \\
q_R(t)&=&- \frac{R(t)\ddot{R}(t)}{\dot{R}^2(t)}=1+2\tan^2(\omega
t),
\nnb \\
H_a(t)&=&\frac{\dot{a}(t)}{a(t)}=-\frac{\omega}{2}(\cot(\omega
t)+2\tan(\omega t)),
\nnb \\
q_a(t)&=&-
\frac{a(t)\ddot{a}(t)}{\dot{a}^2(t)}=-\frac{2\cos^2(\omega
t)(5+\cos(2\omega t))}{(-3+\cos(2\omega t))^2}.
 \eea
\begin{figure}
\vskip 1.5 cm
    \includegraphics{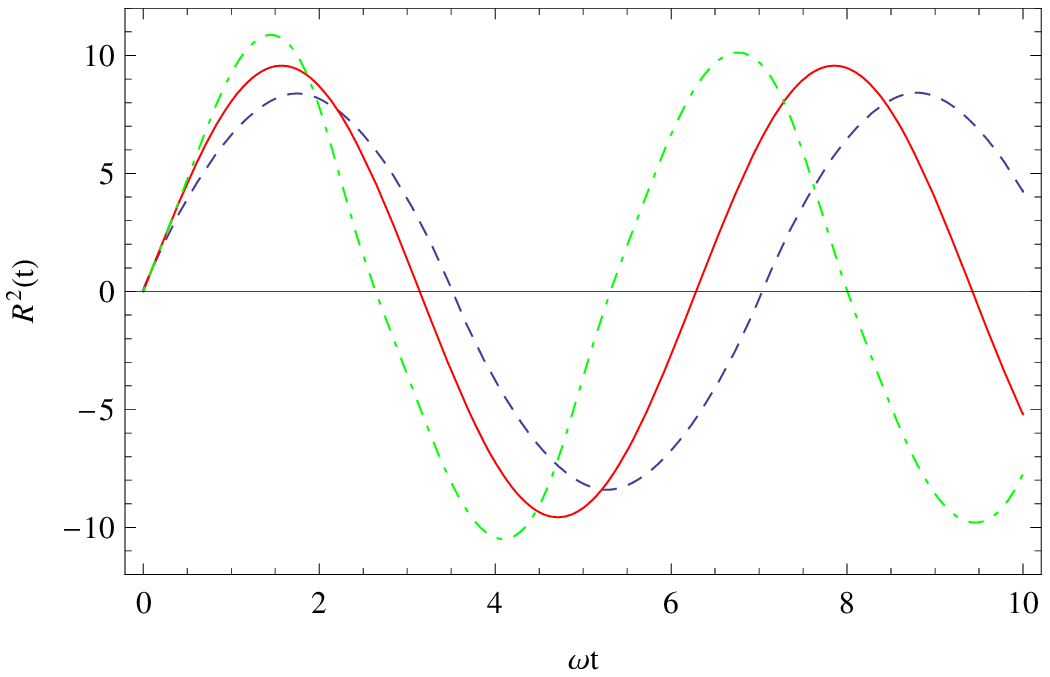}
    \includegraphics{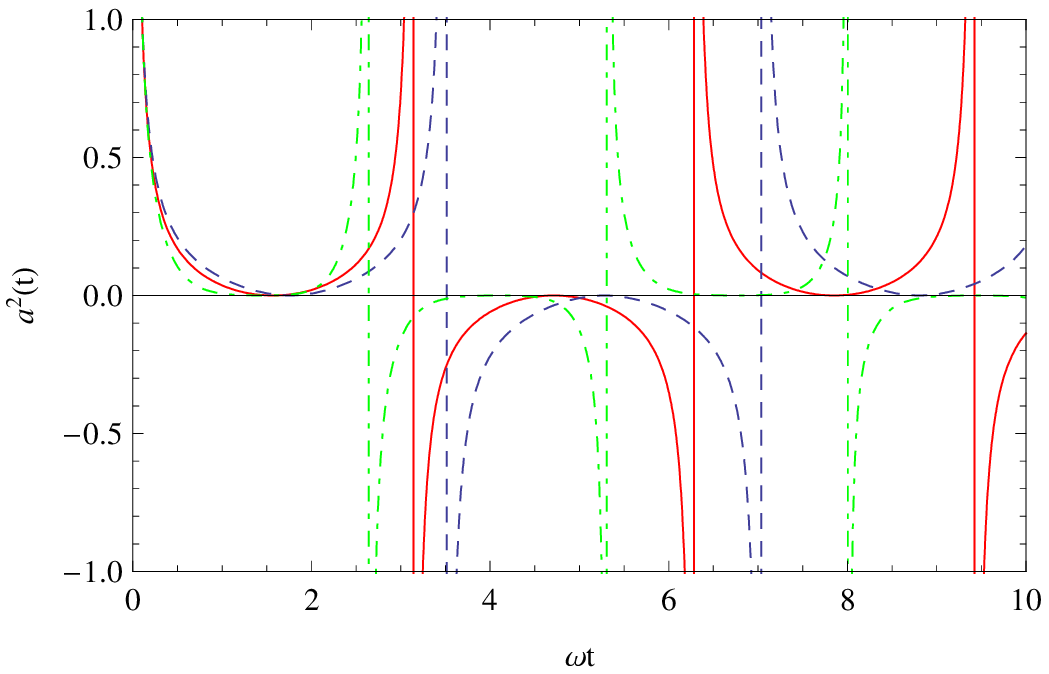}
\vskip 2.5cm \caption{Time evolution of the (squared) scale factors of
Universe with one extra dimension and negative cosmological
constant. Solid, dashed and dot dashed lines refer to the scale factors in commutative, DSR-GUP and GUP framework respectively. Left and
right figures are the external and internal dimensions
respectively. }
\end{figure}
 
The time evolution of $R^2(t)$ and $a^2(t)$ are depicted in Fig.1 (solid lines). According to this behavior, the Universe begins from a big
bang at $t=0$, expands till $t=\frac{\pi}{2\omega}$ toward a
maximum value, and starts contracting toward a big crunch at
$t=\frac{\pi}{\omega}$. As is clear from the figure, as well as the solution $R(t)$ in (\ref{e26}), the big bang is initiated by a anti-de Sitter phase in the case of negative cosmological constant.

Using the present value of Hubble constant, the age of Universe becomes 
$t_{present}=\frac{1}{\omega}\cot^{-1}(\frac{2H_0}{\omega})\approx \omega^{-1}\approx10^{17}s$
which is in agreement with current observations\footnote{The fact that, when choosing the age of the Universe as the inverse of the square root of the cosmological constant (see (\ref{e18})), we end up in the ``most favourable" scenarios (smallest internal scale factor, largest external one) is not so surprising and is just a mere consequence of this choice.}. The present
Universe is also in tideway to get to maximum and minimum of $R^2(t)$
and $a^2(t)$, respectively, within $\Delta t\approx
0.57\omega^{-1}$. 

We set the initial condition at planck time $R(t_{Pl})=a(t_{Pl})$ and according to Fig.1, we see that during the whole time evolution of Universe ($t_{Pl}\leq t\leq
\frac{\pi}{\omega}-t_{Pl})$, the scale factor of internal space is
contracted towards the sizes very smaller than $a(t_{Pl})$, and so can
never exceed $a(t_{Pl})$. Moreover, considering
\bea\label{e29}
R(t_{Pl})=k_2\sqrt{\sin(\omega t_{Pl})},\nnb \\
a(t_{Pl})=k_1\frac{\cos(\omega t_{Pl})}{\sqrt{\sin(\omega
t_{Pl})}},
 \eea
the above initial condition results in
 \bea\label{e30}
\frac{k_2}{k_1}=10^{61}, \eea 
by which we obtain the following ratio

\bea\label{e31} \frac{R(t)}{a(t)}=10^{61}\tan(\omega t). \eea
If the present radius of external space be equal to
the radius of observed Universe $10^{28}cm$, then the present radius of internal space becomes about the Planck length $(10^{-33}cm)$ and this justifies the
non observability of the extra dimension.

For a positive cosmological constant $\omega^2$ is negative, so by replacing $\omega^2$ with $-\omega^2$ in Eq.(\ref{e21}) and using the Hamiltonian constraint the new solutions are obtained
 \bea\label{e32}
R(t)&=&k_2[\cosh(\omega t)]^{\frac{-\rho_-}{\rho_+ \sigma_+
-\rho_- \sigma_-}}[\sinh(\omega t)]^{\frac{\rho_+}{\rho_+ \sigma_+
-\rho_- \sigma_-}},\\
a(t)&=&k_1[\cosh(\omega t)]^{\frac{\sigma_+}{d(\rho_+ \sigma_+ -\rho_- \sigma_-)}}[\sinh(\omega t)]^{\frac{-\sigma_-}{d(\rho_+ \sigma_+ -\rho_- \sigma_-)}}\nnb,
 \eea
where for $d=1$ we have
\bea\label{e33}
a(t)&=&k_1\frac{\cosh(\omega t)}{\sqrt{\sinh(\omega t)}}, \nnb \\
R(t)&=&k_2\sqrt{\sinh(\omega t)}
 \eea
\bea \label{e35} \frac{R(t)}{a(t)}=10^{61}\tanh(\omega t),
 \eea 
and
\bea\label{e36}
H_R(t)&=&\frac{\dot{R}(t)}{R(t)}=\frac{\omega}{2}\coth(\omega t),
\nnb \\
q_R(t)&=&- \frac{R(t)\ddot{R}(t)}{\dot{R}^2(t)}=1-2\tanh^2(\omega
t),
\nnb \\
H_a(t)&=&\frac{\dot{a}(t)}{a(t)}=\frac{\omega}{2}(-\coth(\omega
t)+2\tanh(\omega t)),
\nnb \\
q_a(t)&=&-
\frac{a(t)\ddot{a}(t)}{\dot{a}^2(t)}=-\frac{2\cosh^2(\omega
t)(5+\cosh(2\omega t))}{(-3+\cosh(2\omega t))^2}.
 \eea
\begin{figure}
\vskip 1.5 cm
    \includegraphics{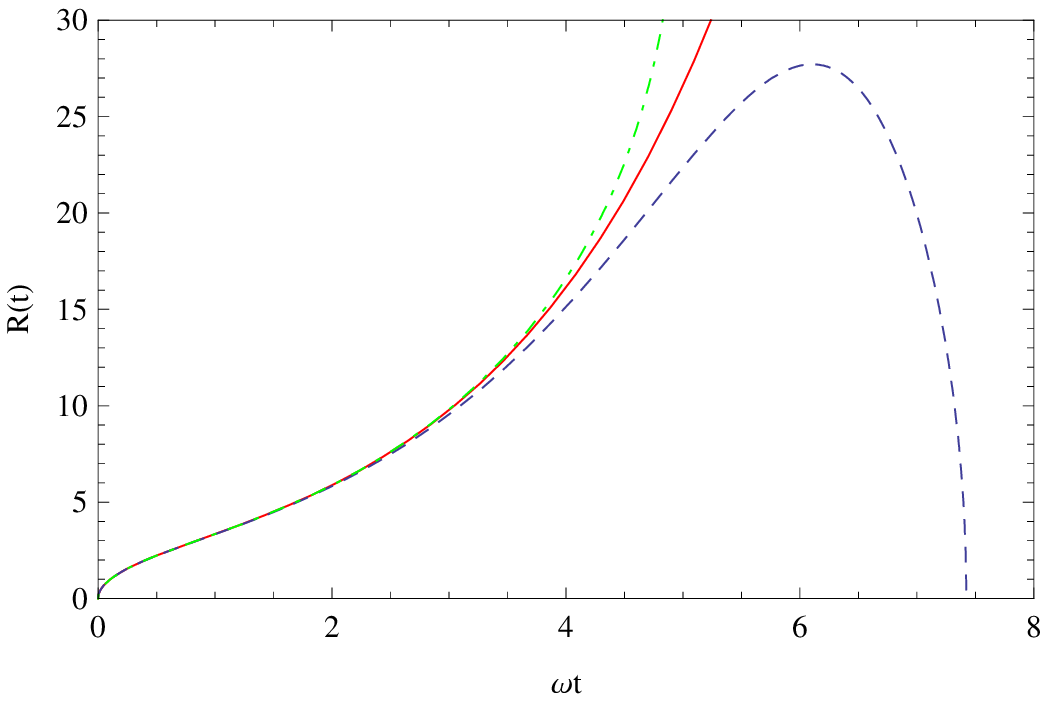}
    \includegraphics{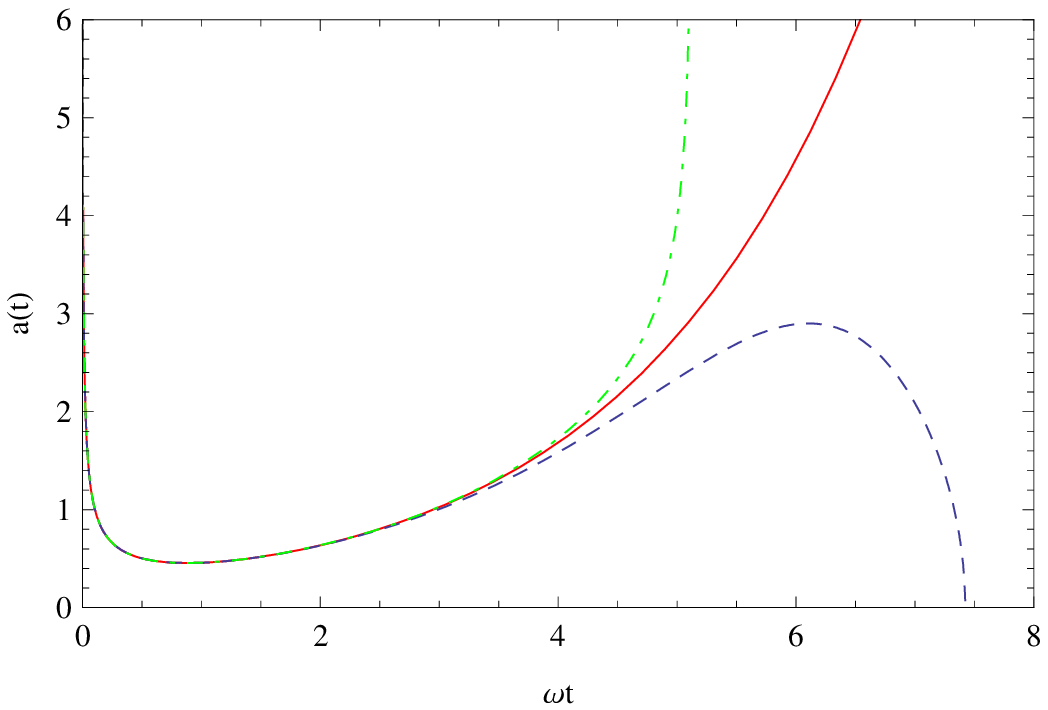}
\vskip 2.5cm \caption{Time evolution of the scale factors of
Universe with one extra dimension and positive cosmological
constant. Solid, dashed and dot dashed lines refer to the scale factors in commutative, DSR-GUP and GUP framework respectively. Left and
right figures are the external and internal dimensions
respectively.}
\end{figure} 

As in the case of negative cosmological constant, the
magnitude of the radius of external to internal spaces is
asymptotically ($t\rightarrow \infty$) about $ 10^{61}$. As is seen in Fig.
2, $R(t)$ is an exponentially increasing function of time whereas $a(t)$ at first decrease with time till $t\simeq0.88\omega^{-1}$ and then increase exponentially. Note that the big bang is initiated by a de Sitter phase in the case of positive cosmological constant.

If the age of Universe is taken as $\omega^{-1}\simeq10^{17}s$, then we find that at present time we are around the minimum point of $a(t)$ and that in the time interval $t_{Pl}\leq t \leq 141\omega^{-1}$, $a(t)$ can never exceeds $a(t_{Pl})$. This indicates that the internal scale factor remains very small, at least for 140 times of the present age of the Universe.

The results obtained here with a positive cosmological constant are consistent with the current observations on the acceleration of the Universe. 
Tho confirm this, we have depicted $H_R$, $H_a$ and $q_R$, $q_a$ in
the figures 3 and 4 (see solid lines). Figure 3 shows that $q_R$
becomes negative a little bit earlier than the present age of the Universe
namely $\omega t \sim1$. This means, the Universe has started its acceleration recently. Fig.4 shows that $q_a$ is always negative and has a minimum at the position where $q_R$ becomes negative. The figures 3 and 4 indicate that at the beginning of time in both commutative and GUP cases, $q_R$ is positive ($R$ is decelerating) and $q_a$ is negative ($a$ is accelerating). With time evolution, $q_R$ approaches the threshold of negative values ($R$ is less decelerating) while $q_a$ approaches to more negative values ($a$ is
highly accelerating). Once $q_R$ enters the region of negative
values ($R$ is accelerating), $q_a$ reaches its minimum ($a$
stops its increasing acceleration). Finally, $q_R$ becomes more negative ($R$ is highly accelerating) whereas $q_a$ goes to rather less negative values ($a$ is slowly accelerating). It is interesting to note that the late time behavior of the Universe is more considerable in the GUP case, where both $R$ and $a$ exhibit highly accelerating features.
\begin{figure}
\vskip 1.5 cm
    \includegraphics{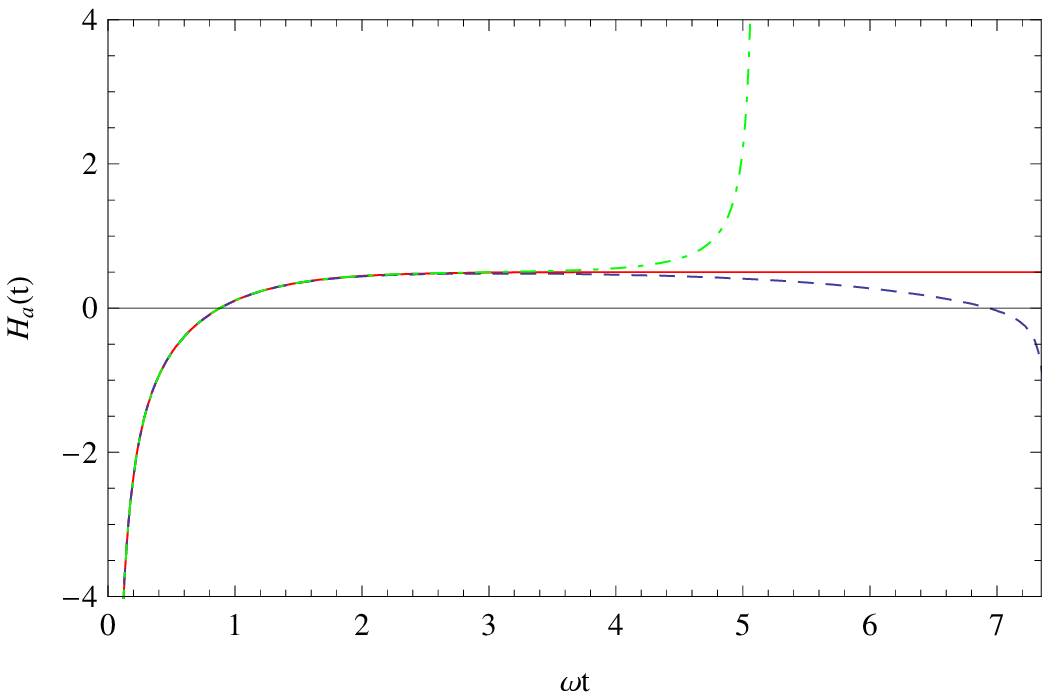}
    \includegraphics{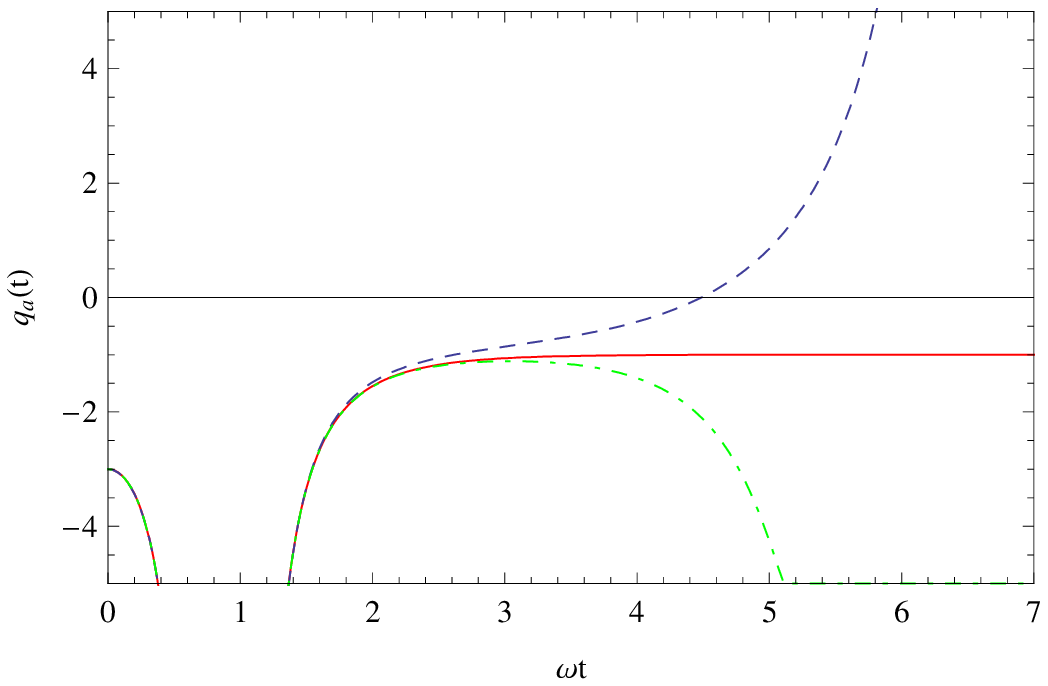}
\vskip 2.5cm \caption{Left and right figures are respectively
Hubble an deceleration parameters of internal space for Universe
with one extra dimension and positive cosmological constant. Solid
, dashed and dot dashed lines refer to the commutative, DSR-GUP and GUP framework
respectively.}
\end{figure}

\subsection{DSR-GUP solutions}

In this section, we aim to study this cosmological model in the DSR-GUP context
to find effects of new terms in commutation relations on the time
evolution of Universe. The new terms in commutation relations can
be considered in two view point: first order in Planck length due
to DSR theory and second order term in Planck length due to GUP in
string theory. Following equation (\ref{e7}) and (\ref{e17}) we
write perturbed Hamiltonian as

 \bea\label{e37}
   {\cal H}=\frac{p_0^2}{2}-\alpha L_{Pl} p_0^3+\frac{(\alpha^2+2\beta)}{2}
   L^2_{Pl}p^4_0+\omega^2(x_1^2-
   x_2^2),
  \eea
where $p^2_0=\frac{p_{10}^2}{2}-\frac{p_{20}^2}{2}$ and
$[x_{i0},p_{j0}]=i \delta_{ij}$. Here, we want to investigate the
classical version of DSR-GUP algebra. To do this, we must replace
the quantum mechanical commutators with the classical poisson
bracket as $[P,Q]\rightarrow i\{P,Q\}$. Using equation (
\ref{e19}), the equations of motion can be written as

\bea\label{e38}
\dot{x}_\mu&=&\{x_\mu,{\cal H}\}_P=\frac{1}{2}p_\mu-\frac{3}{2}\alpha L_{Pl}p_0p_\mu+(\alpha^2+2\beta)L^2_{Pl}p^2_0p_\mu,\nnb \\
\dot{p}_\mu&=&\{p_\mu,{\cal H}\}_P=-2\omega^2x_\mu.
\eea
\begin{figure}
\vskip 2.5 cm
    \includegraphics{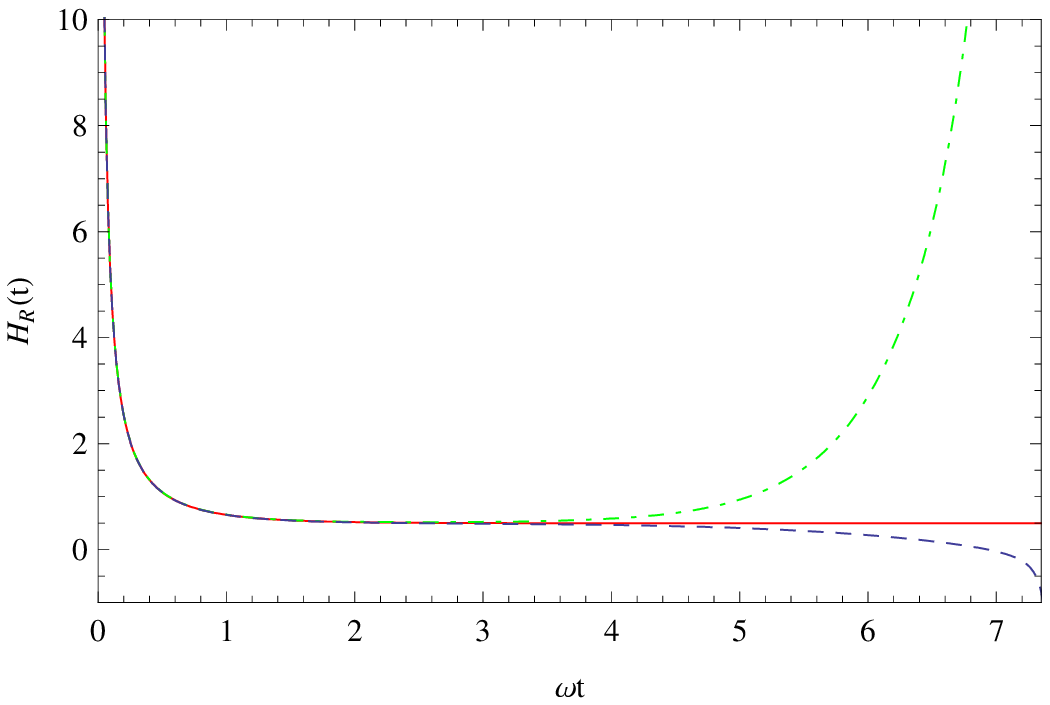}
    \includegraphics{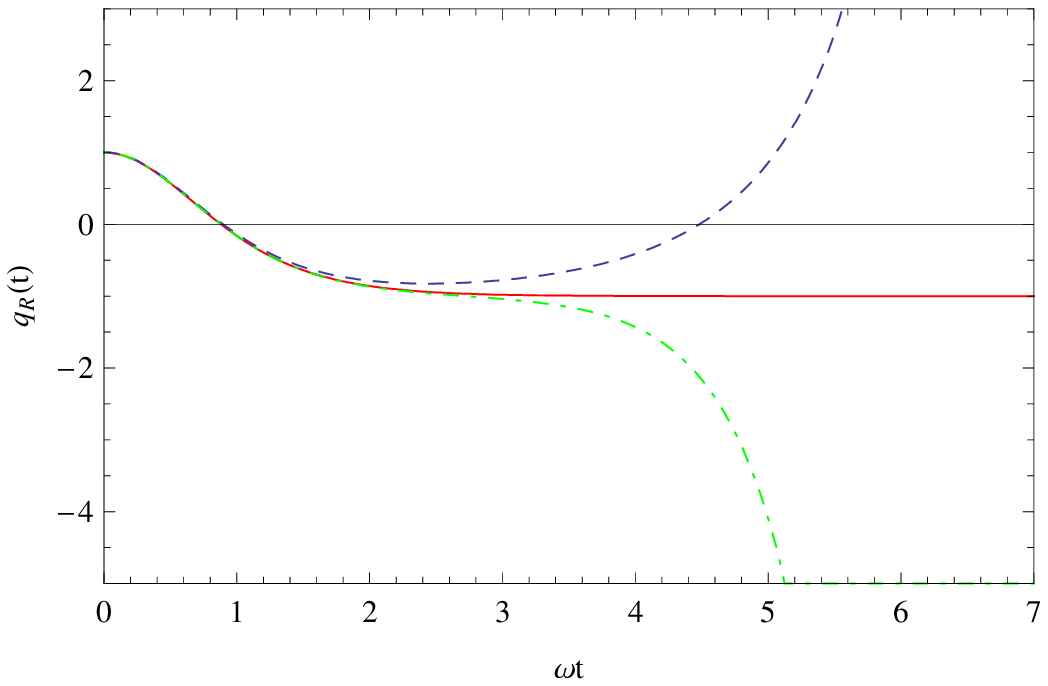}
\vskip 2.5cm \caption{Left and right figures are respectively
Hubble an deceleration parameters of external space for Universe
with one extra dimension and positive cosmological constant. Solid
, dashed and dot dashed lines refer to the commutative, DSR-GUP and GUP framework
respectively.}
\end{figure}
We see that deformed classical equations form a system of nonlinear
coupled differential equation, so we need numerical solutions.
Setting $\alpha=0$ reduces the equations to GUP framework so we can
see effect of the second order term in Planck length on the
time evolution of Universe. We limit ourselves to the investigation of the effect of first order term in Planck length due to DSR theory. In so doing, we should set $\beta= 0$ and $\alpha=\epsilon$, $\epsilon$ being a nonvanishing
but so small parameter that $\alpha^2\simeq 0$. This removes the third term,
second order in Planck length, in the right hand side of the first equation
in (\ref{e38}).

In the negative cosmological constant framework, $\omega^2$ is
positive. Numerical solution of equations (\ref{e38}) shows that
the deformed scale factors like commutative and GUP cases have periodic
behavior. To compare with the results obtained in GUP \cite{zeynali}, we have included the behaviour of the scale factors in GUP as well as commutative cases within the figures. As is seen in Fig.1, the time interval between big bang and big crunch in GUP case is shortened with respect to commutative one while in DSR-GUP case the time interval between big bang and big crunch is longer. In GUP case, the deformed scale factor of the internal space reaches it's minimum value sooner than commutative one while the deformed scale factor of the internal space in DSR-GUP case reaches later
than commutative one. The deformed scale factor of the external space in GUP case reaches it's maximum sooner than commutative one and has larger value while in the DSR-GUP case, the deformed scale factor of the external space has smaller value and reaches to it later than commutative one.

Replacing $\omega^2$ with $-\omega^2$ in Eq.(\ref{e38}),
leads to the corresponding equations in the case of positive
cosmological constant. Numerical analysis shows that (Fig.2) at early times
the deformed scale factors of internal and the external spaces in both 
GUP and DSR-GUP cases behave like commutative one, and specifically the Universe
is initiated by a de Sitter phase.  At later times in the GUP framework, the expanding rate of deformed scale factors are bigger than commutative case while in the DSR-GUP case the deformed scale factors increase
slower than commutative case. Moreover, at very late times after experiencing a maximum value, both scale factors decrease towards a big crunch simultaneously.
Looking at the behaviour of deceleration parameter in figures 3 and 4 for
internal and external scale factors, respectively, shows that unlike the
commutative and GUP cases, in DSR-GUP case both scale factors after accelerating phases experience a decelerating phase at late times. This makes a remarkable
difference between the DSR-GUP in one hand, and commutative together with GUP cases on the other hand. The difference between GUP and DSR-GUP is due
to a relative sign difference in the algebras (\ref{e2}) and (\ref{e3}) corresponding to GUP and DSR-GUP. This is interesting because the final fate of our multidimensional
cosmology, being accelerated forever or decelerated towards a big crunch,
is simply related to a relative sign difference in the quantum algebra corresponding
to two different generalized uncertainty principles, GUP and DSR-GUP. Note
also that the maximums of scale factors and the temporal location of
big crunch in DSR-GUP case (Fig.2) depends on the Planck length: the more
smaller Planck length, the more distant big crunch.

\section{Discussion and Conclusion}

We have studied a multidimensional cosmology having FRW type metric with a 4-dimensional space-time sector and a $d$-dimensional Ricci-flat internal space subjected to a higher dimensional cosmological constant in the frameworks
of commutative and DSR-GUP contexts. The corresponding exact solutions for negative and positive cosmological constants are obtained and compared with
each other as well as GUP case. It is shown in DSR-GUP case that for positive cosmological constant, both scale factors of internal and external spaces
after accelerating phase, unlike the commutative and GUP cases, will inevitably experience decelerating phase leading simultaneously to a big crunch. This unexpected behaviour originates simply from a negative sign in the DSR-GUP algebra. The important result in this model is that DSR-GUP prevents the
eternal acceleration. 

The exact solutions which we have obtained are the background cosmologies.
Although they are interesting in themselves, but it is useful to provide insight on the limits of our approach, and the extent of viability of these solutions. The background cosmological solutions obtained here describe an ideal picture of the Universe and its evolution subject to Ricci flat extra dimensions, generalized uncertainty principle and positive/negative cosmological
constant. No real matter is assumed, however, it is possible to interpret the contribution of extra dimensions as a kind of effective matter. Actually,
the presence of real matter complicates the calculations, hence our approach is limited to the vacuum with a cosmological constant. We have also limited ourselves to the investigation of the effect of first order term in Planck length which affects the equations of motion due to DSR-GUP algebra. Note, however, that using the generalized uncertainty principle, including the Planck length in the corresponding algebra, does not mean that we have involved directly with quantum gravity. 

In this paper, we did not pay attention to the issue of perturbations, however
it is useful to briefly mention about this subject. In fact, a realistic description of Universe needs the study of homogeneous and isotropic perturbations and the stability of these solutions against the perturbations. In this regard, we may use a set of convenient phase-space variables and write the Friedmann equation in terms of dimensionless density parameters. Then, we obtain a system of differential equations. The behavior of this system of equations in the neighborhood of its stationary point is determined by the corresponding matrix of its linearization. The real parts of its eigenvalues tell us whether the corresponding cosmological solution is stable or unstable with respect to the homogeneous and isotropic
perturbations. If the real part of the eigenvalues of a critical point is not zero, the point is said to be hyperbolic. In this case, the dynamical character of the critical point is determined by the sign of the real part of the eigenvalues. If all of them are positive, the point is said to be a {\it repeller}, because arbitrarily small deviations from this point will move the system away from this state. If all of them are negative, the point is called an {\it attractor} because if we move the system slightly from this point in an arbitrary way, it will return to it. Otherwise, we say the critical point is a {\it saddle point}.

\section*{Acknowledgment}
The authors would like to thank the anonymous referee for the enlightening comments.

\end{document}